\documentclass[12pt]{article}
\usepackage{graphicx}
\usepackage{epsfig}
\usepackage{amsmath}
\usepackage{amsfonts}
\usepackage{amssymb}
\def \ptrel {$P_{T}^{\,\rm rel}~$}
\def \gm {$\gamma+\mu~$}

\begin{document}
\begin{titlepage}

\hfill
\begin{tabular}{r}
FERMILAB-Pub-01/087-E \\
\end{tabular}

\vspace{0.5cm}

\begin{center}


{\LARGE Cross Section and Heavy Quark Composition of \\ $\gamma+ \mu
$ Events Produced in $p \bar p$ Collisions}

\vspace{0.25cm}

\font\eightit=cmti8
\def\r#1{\ignorespaces $^{#1}$}
\hfilneg
\begin{sloppypar}
\noindent
T.~Affolder,\r {23} H.~Akimoto,\r {45}
A.~Akopian,\r {38} M.~G.~Albrow,\r {11} P.~Amaral,\r 8  
D.~Amidei,\r {26} K.~Anikeev,\r {24} J.~Antos,\r 1 
G.~Apollinari,\r {11} T.~Arisawa,\r {45}
% ...
\end{sloppypar}
\vskip .026in
\begin{center}
(CDF Collaboration)
\end{center}

\vskip .026in
\begin{center}
\r 1  {\eightit Institute of Physics, Academia Sinica, Taipei, Taiwan 11529, 
Republic of China} \\
\r 2  {\eightit Argonne National Laboratory, Argonne, Illinois 60439} \\
\r 3  {\eightit Istituto Nazionale di Fisica Nucleare, University of Bologna,
I-40127 Bologna, Italy} \\
\r 4  {\eightit Brandeis University, Waltham, Massachusetts 02254} \\
\r 5  {\eightit University of California at Davis, Davis, California  95616} \\
\r 6  {\eightit University of California at Los Angeles, Los 
Angeles, California  90024} \\  
\r 7  {\eightit Instituto de Fisica de Cantabria, CSIC-University of Cantabria, 
39005 Santander, Spain} \\
\r 8  {\eightit Enrico Fermi Institute, University of Chicago, Chicago, 
Illinois 60637} \\
\r 9  {\eightit Joint Institute for Nuclear Research, RU-141980 Dubna, Russia}
\\
\r {10} {\eightit Duke University, Durham, North Carolina  27708} \\
\r {11} {\eightit Fermi National Accelerator Laboratory, Batavia, Illinois 
60510} \\
\r {12} {\eightit University of Florida, Gainesville, Florida  32611} \\
\r {13} {\eightit Laboratori Nazionali di Frascati, Istituto Nazionale di Fisica
               Nucleare, I-00044 Frascati, Italy} \\
\r {14} {\eightit University of Geneva, CH-1211 Geneva 4, Switzerland} \\
\r {15} {\eightit Glasgow University, Glasgow G12 8QQ, United Kingdom}\\
\r {16} {\eightit Harvard University, Cambridge, Massachusetts 02138} \\
\r {17} {\eightit Hiroshima University, Higashi-Hiroshima 724, Japan} \\
\r {18} {\eightit University of Illinois, Urbana, Illinois 61801} \\
\r {19} {\eightit The Johns Hopkins University, Baltimore, Maryland 21218} \\
\r {20} {\eightit Institut f\"{u}r Experimentelle Kernphysik, 
Universit\"{a}t Karlsruhe, 76128 Karlsruhe, Germany} \\
\r {21} {\eightit Center for High Energy Physics: Kyungpook National
University, Taegu 702-701; Seoul National University, Seoul 151-742; and
SungKyunKwan University, Suwon 440-746; Korea} \\
\r {22} {\eightit High Energy Accelerator Research Organization (KEK), Tsukuba, 
Ibaraki 305, Japan} \\
\r {23} {\eightit Ernest Orlando Lawrence Berkeley National Laboratory, 
Berkeley, California 94720} \\
\r {24} {\eightit Massachusetts Institute of Technology, Cambridge,
Massachusetts  02139} \\   
\r {25} {\eightit Institute of Particle Physics: McGill University, Montreal 
H3A 2T8; and University of Toronto, Toronto M5S 1A7; Canada} \\
\r {26} {\eightit University of Michigan, Ann Arbor, Michigan 48109} \\
\r {27} {\eightit Michigan State University, East Lansing, Michigan  48824} \\
\r {28} {\eightit University of New Mexico, Albuquerque, New Mexico 87131} \\
\r {29} {\eightit The Ohio State University, Columbus, Ohio  43210} \\
\r {30} {\eightit Osaka City University, Osaka 588, Japan} \\
\r {31} {\eightit University of Oxford, Oxford OX1 3RH, United Kingdom} \\
\r {32} {\eightit Universita di Padova, Istituto Nazionale di Fisica 
          Nucleare, Sezione di Padova, I-35131 Padova, Italy} \\
\r {33} {\eightit University of Pennsylvania, Philadelphia, 
        Pennsylvania 19104} \\   
\r {34} {\eightit Istituto Nazionale di Fisica Nucleare, University and Scuola
               Normale Superiore of Pisa, I-56100 Pisa, Italy} \\
\r {35} {\eightit University of Pittsburgh, Pittsburgh, Pennsylvania 15260} \\
\r {36} {\eightit Purdue University, West Lafayette, Indiana 47907} \\
\r {37} {\eightit University of Rochester, Rochester, New York 14627} \\
\r {38} {\eightit Rockefeller University, New York, New York 10021} \\
\r {39} {\eightit Rutgers University, Piscataway, New Jersey 08855} \\
\r {40} {\eightit Texas A\&M University, College Station, Texas 77843} \\
\r {41} {\eightit Texas Tech University, Lubbock, Texas 79409} \\
\r {42} {\eightit Istituto Nazionale di Fisica Nucleare, University of Trieste/
Udine, Italy} \\
\r {43} {\eightit University of Tsukuba, Tsukuba, Ibaraki 305, Japan} \\
\r {44} {\eightit Tufts University, Medford, Massachusetts 02155} \\
\r {45} {\eightit Waseda University, Tokyo 169, Japan} \\
\r {46} {\eightit University of Wisconsin, Madison, Wisconsin 53706} \\
\r {47} {\eightit Yale University, New Haven, Connecticut 06520} \\
\r {(\ast)} {\eightit Now at Carnegie Mellon University, Pittsburgh,
Pennsylvania  15213} \\
\r {(\ast\ast)} {\eightit Now at Northwestern University, Evanston, Illinois 
60208}
\end{center}

\newpage

\begin{abstract}
We present a measurement of the cross section and the first measurement of the heavy flavor content of associated direct photon
+ muon events produced in hadronic collisions.  These measurements come from a sample of 1.8~TeV $p \bar
p$ collisions recorded with the Collider Detector at Fermilab.
Quantum chromodynamics (QCD) predicts that these events are primarily due
to Compton scattering process $cg\rightarrow c\gamma$,
with the final state charm quark producing a muon.
The cross section for events with a photon transverse momentum between 12 and 40 GeV/c is measured to be $46.8 \pm
6.3 \pm
7.5$ pb,  which is two standard deviations below the most recent theoretical calculation.
A significant fraction of the events in the sample contain a final-state bottom quark.  The ratio of charm to bottom production is measured to be $2.4 \pm
1.2$, in good agreement with QCD models.
\end{abstract}
\end{center}
\vfill
\newpage
\end{titlepage}

Measurements of the inclusive spectrum of direct photons
in hadron-hadron collisions have provided important 
tests of quantum chromodynamics (QCD).
Similar tests have been made with inclusive measurements of heavy
flavor production (b and c quarks).  The data and current perturbative
QCD models do not agree well for both inclusive processes, giving
insights into possible limitations of such
models\cite{photon}\cite{hqreview}.  Two previous measurements of the
\emph{associated} production of direct photons and charm quarks
have provided checks of the charm quark content of the
proton\cite{dstar}\cite{phomu1a} through the Compton scattering
process $cg\rightarrow c\gamma$.  We present here an analysis with
an order-of-magnitude more events, collected by the Collider Detector at
Fermilab (CDF), that provides a quantitative test of perturbative
QCD.  In addition, this new measurement is sensitive
to the production of bottom quarks in association with the photon.

%
%
%
The associated production of direct photons and heavy quarks in hadron 
collisions is expected to be a unique system for the study of the charm quark,
 with a 9:1 ratio of charm to bottom quarks in parton level QCD\ calculations\cite{nloqcd}\cite{pythia}.
\ Typically in hadron collisions, heavy quarks are produced in the gluon-gluon
initiated processes $gg\rightarrow Q\overline{Q}$ and $gg\rightarrow
gg\rightarrow gQ\overline{Q}$ \ where one of the final state gluons splits
into the heavy quark pair. \ In either case\ if the gluon energy is
sufficiently larger than the bottom quark mass, \ the production of bottom pairs is
approximately equal to that of charm pairs. \ \ In semi-leptonic decays of
heavy quarks, \ the harder fragmentation function of the bottom quark leads to
its dominance in these samples (for example a 1:4 ratio of charm to bottom in
reference\cite{inclusivemu}). \ \ \ The direct-photon Compton process,
however, is proportional to the quark electric charge squared, which increases
the ratio of charm to bottom by a factor of 4. \ In addition, the 
intrinsic bottom quark content in the proton is 60\% smaller than the charm 
quark content in our kinematic region. The combination of the quark charge
coupling and the different proton content means the charm quark is
expected to play the larger role in direct photon events. \ In this Letter we
present the first measurement of the charm and bottom composition of direct
photon events in hadronic interactions.

%
%
%
The data for this analysis are from an integrated luminosity of 86~pb$^{-1}$
of $p\bar{p}$ collisions collected with CDF in the 1994-95 Tevatron collider
run (Run 1b). The CDF detector and its coordinate system has been described in
detail elsewhere\cite{detcdf}\cite{top}. The events in the photon data sample
discussed in this paper triggered the experiment by satisfying the requirement
of a photon and a muon candidate at the hardware trigger level, \ whereas in
the previous measurements only the photon candidate was required by the
trigger. \ This allowed a lower transverse momentum $P_{T}(=Psin(\theta))$
threshold, \ in this case 10 GeV. \ A\ photon candidate is selected by
requiring a cluster of energy in the central electromagnetic calorimeter
$|\eta_{\gamma}|<0.9$, with no charged tracks pointing to the cluster. The
clusters are required to have a photon $P_{T}$ between 12 and 40~GeV and to be
isolated, with less than 1 GeV of additional transverse energy in a cone of
$\Delta R=\sqrt{\Delta\phi^{2}+\Delta\eta^{2}}=0.4$ around the cluster.
\ Additional photon cuts were used which were identical to those used in the
Run 1a CDF inclusive photon analysis\cite{cdfpho}. \ Muon candidates were
selected by requiring a match between a charged track with $P_{T}>4$ GeV/c in
the central tracking chamber and a track in the appropriate muon system. \ For
$|\eta_{\mu}|<0.6$ muon candidates are required
to be identified in both the CMU and in the Central Muon Upgrade
system (CMP), which is behind an additional 1-m thickness of
steel. For $0.6<|\eta _{\mu}|<1.0$ the muon candidate track was
required to be reconstructed in the central muon extension system
(CMX).  All three muon systems are discussed in detail in
reference\cite{phomu1a}\cite{thesis}.  After the track matching
requirement there are 3850 events with a direct photon candidate and
a muon candidate.

%
%
%
Photon backgrounds from $\pi^{0}$ and $\eta$ meson decays remain in the
sample. \ They are subtracted on a statistical basis using the photon
background subtraction ``profile'' method described in reference\cite{cdfpho}.
\ This method uses the transverse energy profile of the electromagnetic shower
as a discriminant between single direct photons and multiple-photon meson
decays. \ The requirement of photon $P_{T}<40$ GeV/c described above is
necessary in order to use this technique. \ After subtracting these
backgrounds, 1707$\pm83$ direct photons with a muon candidate remain. \ For
comparison, the previous publication\cite{phomu1a} was based on 140 events. \ 

%
%
%
Muon backgrounds from charged pion and kaon decays remain in this sample, \ as
well as a smaller fraction of charged hadrons that do not interact
significantly in the material in front of the muon detectors. \ \ These are
estimated with the same technique as in the previous analysis. \ Starting with
the parent inclusive photon + jet data sample, \ the 4-vector of each charged
particle with $P_{T}>0.4$ GeV/c is measured. \ Each track is passed
into a detector simulation as a charged pion or kaon, with a $\pi/K$
ratio of 60\%/20\%\cite{e731}. The results of the simulation are
passed through the muon reconstruction; the tracks passing all cuts
form the sample used for the background estimate.  Backgrounds from
protons that penetrate the calorimeter are negligible.  After
statistically subtracting the muon backgrounds, we expect 724$\pm89$
direct photon events with a muon that is not from charged $\pi^{\pm}$
or $K^{\pm}$ decay.  These events are assumed to come from associated
direct photon + heavy quark production with the heavy quark decaying
into a muon.  Figure 1 shows the number of signal and background
events in five bins of photon P$_{T}$.  Note that the purity of the
sample improves dramatically as the photon P$_{T}$ increases, which is
due to the improved rejection of neutral meson backgrounds. The
measured purity is consistent with that in inclusive direct photon
measurements\cite{cdfpho}.

%
%
%
The photon-muon cross section $d\sigma^{\gamma+\mu}/dP_{T}^{\gamma}$ is
derived for these five bins in photon P$_{T}$ by dividing by the luminosity,
86 $pb^{-1}$, the photon P$_{T}$ bin size in GeV/c, and the efficiencies for
detecting the photon within $|\eta_{\gamma}|<0.9$ and the muon within
$|\eta_{\mu}|<1.0.$ \ These efficiencies include the detector acceptance
within the relevant pseudorapidity range, \ but are defined after the photon
or muon P$_{T}$ cut. \ The efficiencies are measured by a combination of
studies using Monte Carlo simulation and data\cite{phomu1a}\cite{thesis}.
\ The photon efficiency varies from $34\%$ to $38\%$ with a small photon
$P_{T}$ dependence, while the muon efficiency varies from $49\%$ to $53\%$ and
depends slightly on the specific muon subsystem. \ \ The resulting photon-muon
cross section is shown in Table 1 along with the statistical uncertainties.

%
%
%
There are four significant systematic uncertainties on the direct photon +
muon cross section: 1) 12\% from the muon background subtraction, which mostly
comes from the uncertainties in the estimated pion and kaon particle
fractions; 2) 7\% from the photon background subtraction uncertainty,
\ estimated in the inclusive photon measurement;  3) 7\% from the uncertainty
in the photon and muon cut efficiencies; \  and 4) 4.3\% from the uncertainty
in the CDF luminosity measurement, \ which is predominantly due to the
uncertainty in the total $\overline{p}p$ cross section. These added in
quadrature give an uncertainty that ranges from 16\% to 20\% as the photon P$_{T}$ increases from 12 to 40 GeV. \ 

%
%
%
\begin{table}[ptb]
\begin{minipage}[h]{6.5in}
\begin{center}
\small{
\begin{tabular}
[c]{|c|c|c|c|}\hline
Photon P$_{T}$ Bin (GeV/c) & $d\sigma^{\gamma+\mu}/dP_{T}^{\gamma}$ (pb/GeV/c) & PYTHIA (pb/GeV/c) &
NLO\ QCD (pb/GeV/c)\\\hline
12-14 & 7.5$\pm1.9$ & 3.4 & 10.3\\\hline
14-17 & 4.4$\pm1.0$ & 2.3 & 6.4\\\hline
17-20 & 1.7$\pm0.7$ & 1.5 & 3.8\\\hline
20-26 & 1.5$\pm0.4$ & 0.9 & 1.9\\\hline
26-40 & 0.3$\pm0.2$ & 0.3 & 0.5\\\hline
\end{tabular}
\caption
{The measured photon-muon cross section and the predictions from PYTHIA and NLO QCD are tabulated in five bins of photon
P$_{T}$.}
}
\end{center}
\end{minipage}
\end{table}

The photon-muon cross section is compared to two different QCD
calculations of photon-muon production. The first calculation is that
in the PYTHIA~\cite{pythia} Monte Carlo, which only has the
leading-order (LO) contributions to the photon+heavy quark cross
section, but has the full parton shower and fragmentation effects.
The CLEO heavy quark decay tables are used\cite{cleomc}.  The second
calculation is a next-to-leading order (NLO) QCD photon+heavy flavor
calculation~\cite{nloqcd}, which has additional processes not present
at leading-order, of which $gg\rightarrow c\overline{c}\rightarrow
c\overline{c}\gamma$ is the largest contributor, but which operates
only at the parton level and thus does not include heavy quark
fragmentation. For this we use the Peterson fragmentation model in
PYTHIA and the CLEO decay tables.  The probability of observing a 4
GeV muon from this model of the decay is shown in Fig.~2 as a function
of heavy quark $P_{T}$ for both direct bottom and charm decays and for
sequential decays of the bottom quark.  The NLO QCD calculation uses
the massless quark approximation, which is adequate for photon+charm
since the scale of the process is well above the charm mass, but
cannot be applied to the photon+bottom process.  The photon+bottom
part of this calculation is derived from a leading-order photon+bottom
calculation including mass effects, to which a multiplicative factor
(`K factor', {\it K}) is then applied to account for higher-order
effects. We use {\it K}$=1.8$, which is the calculated ratio of NLO/LO
in inclusive bottom production in a kinematic range that is close to
that of the current measurement\cite{kfactor}.  The uncertainties in
the NLO calculation have not been thoroughly studied, but it is likely
the K factor is the largest component.  If one assigns a 50\% error to
{\it K}, then the predicted total photon+heavy flavor NLO cross
section has a 12\% uncertainty.  Variations of renormalization scale,
parton distributions, and fragmentation functions all give variations
of $\approx5\%.$

%
%
%
The NLO QCD cross section, as well as the PYTHIA predictions, are compared to
the data in Table 1 and Figure 3. \ The NLO cross sections are much
larger than those calculated with PYTHIA, \ due to the inclusion of additional
processes mentioned above. \ The shape of the data shown in Figure 3 matches
both calculations, \ while the normalization is a factor 1.9 larger than the
PYTHIA prediction and a factor 1.45 smaller than predicted by NLO\ QCD. \ With
a data normalization uncertainty of 16\% and a 12\% uncertainty in the NLO QCD
calculation, \ the data lie about two standard deviations below the NLO QCD
prediction. \ The shape of the photon+muon cross section, however, is better
described by theory than is the inclusive photon $P_{T}$ spectrum. \ In addition,
measured inclusive heavy flavor cross sections have typically been a factor of
2 larger than NLO QCD predictions\cite{hqreview}; we do not observe this in
photon+heavy flavor production. \ 

%
%
%
The current data sample is large enough to study the ratio of charm quark to
bottom quark production in association with the photon. \ A ``jet'' of charged
particles is measured by using the muon candidate as a seed, and then
clustering charged tracks in a cone of radius 0.7 in $\eta-\phi$ space around
the muon. \ We use the transverse momentum of the muon with respect to the jet
axis, P$_{T}^{rel},$ as the variable to separate the charm and bottom
fractions of the sample. \ \ The larger bottom quark mass leads to an enhanced
P$_{T}^{rel}.$ \ A maximum likelihood fit is performed to the distribution of
P$_{T}^{rel}$ from the photon + muon candidate sample to four template
spectra: 1) photon+charm, 2) photon+bottom, 3) photon + fake muon, and 4)
$\pi^{0}$ + X where X is a muon candidate from any source (charm, bottom, or
fake muon). \ \ \ The first two templates are generated using the PYTHIA Monte
Carlo\cite{pythia}. \ \ The model is checked by returning to the parent
photon+jet sample and comparing the P$_{T}^{rel}$ distribution of the highest
P$_{T}$ track in the jet to the same from PYTHIA. \ The modeling is quite good
using default PYTHIA parameters. \ The third template, \ photon + fake muon,
\ comes from the same data+simulation combination used to estimate the fake
muons for the cross section. \ Its normalization in the fit is constrained
with a Gaussian weight using the systematic uncertainty of 12\% coming from
the cross section measurement. \ The fourth template, \ $\pi^{0}$ + X, \ comes
directly from the data as it is the component subtracted during the direct
photon background subtraction. \ \ The systematics on the maximum likelihood
fit using these four templates come from uncertainties in the shape of each
template. \ The systematic uncertainty on the photon+charm quark and
photon+bottom quark templates are estimated by varying the input PYTHIA parameters.   The allowed range of the parameters was determined by comparing the P$_{T}^{rel}$ distribution in the parent
photon+jet data sample with PYTHIA.  This leads to an uncertainty of 0.2 in the
charm/bottom ratio. \ The systematic uncertainty on the two data-driven
templates, \ photon + fake muon and $\pi^{0}$ + X, \ are dominated by the
photon subtraction method uncertainty and is 0.15 in the charm/bottom ratio.
The statistical uncertainties from the maximum likelihood fit are much larger
than the systematic uncertainties, being 1.1 in the charm/bottom ratio. \ \ The
result of the fit to the charm/bottom ratio of the sample is 2.4$\pm1.2$
(stat. + sys.)$,$ to be compared with the predictions of 2.9 by PYTHIA and 3.2
by NLO\ QCD. \ \ \ There is good agreement between data and theory in this
ratio. \ \ The unique nature of this charm-enriched sample is confirmed, \ as
the measured charm/bottom ratio of 2.4 is much larger than the value of
$\ 0.2$ in the inclusive heavy flavor samples in Reference\cite{inclusivemu}.
\ \ The photon+muon P$_{T}^{rel}$ distribution after all events with either a
fake photon or muon are subtracted is shown in Figure 4a, \ compared to the
templates for photon+charm and photon+bottom. \ The templates are normalized to the data. \ \ The
photon+muon data are clearly a combination of charm and bottom, with a
distribution more like the softer charm distribution. \ For comparison the
same distribution using the $\pi^{0}$ + muon events from the same sample is
shown in Figure 4b. \ Although QCD predictions for this process do not exist,
\ qualitatively one would expect a larger bottom content since this process
would not have the matrix-element enhancement due to the square of the 
electric charge of the quark, as does the photon+muon `Compton' process. 
The data confirm this hypothesis, albeit with limited statistics.

%
%
%
In summary,  we present a measurement of direct photon plus
associated muon production in hadronic interactions with an
order-of-magnitude more events than  the previous measurement.
The measurement is an interesting combination of direct photon and heavy
flavor physics,  each of which has had difficulties in comparisons
with NLO QCD calculations.  The data agree in shape with the
theoretical predictions,  but
fall below the theory in normalization by 2 standard deviations.  The ratio
of charm/bottom in the sample has been measured for the first time,  and
confirms the QCD expectation that the sample is very enriched in charm quarks
compared to inclusive lepton samples in hadron collisions.   

%
%
%
\bigskip
We thank the Fermilab staff and the technical staffs of the participating
institutions for their vital contributions. This work was supported by the
U.S. Department of Energy and National Science Foundation, the Italian
Istituto Nazionale di Fisica Nucleare, the Ministry of Science, Culture and
Education of Japan, and the Alfred P. Sloan Foundation.

%
%
%

\bigskip
\begin{figure}[ptbh]
\begin{center}
\begin{minipage}[h]{4.0in}
\epsfig{file=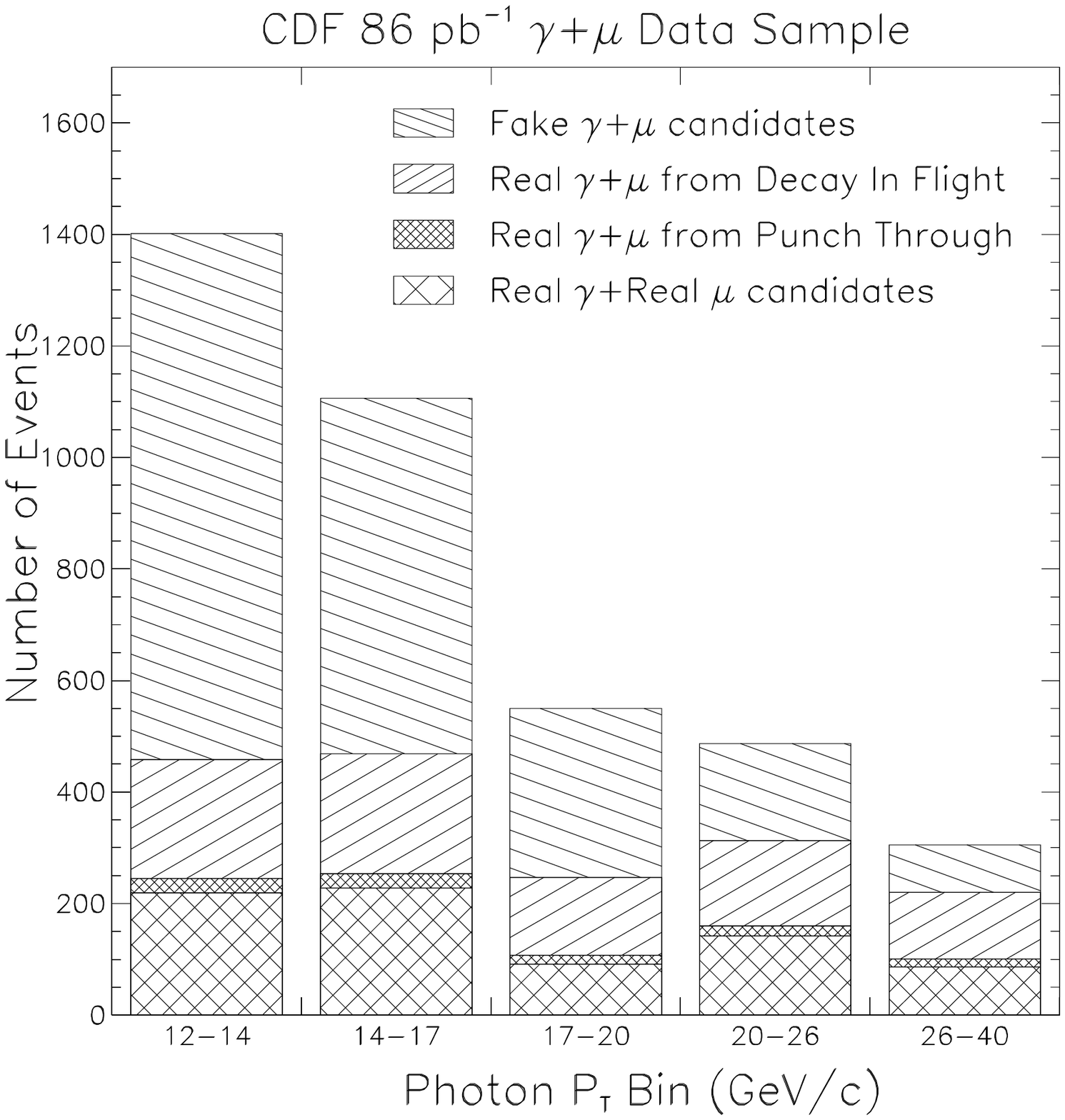,width=\linewidth}\caption
{The number of signal and background events in each bin of photon $P_T$ are
shown for the \gm
sample.  In addition to the signal of a real photon and a muon from a heavy-quark decay, the three components of the background are shown: 1) fake photons plus
real or fake muons, 2) real photons plus a muon coming from the decay of a
charged pion or kaon, and 3) real photons plus a fake muon coming from the
punchthrough of a charged pion or kaon.}%
\end{minipage}
\end{center}
\label{phomunev}\end{figure}
\begin{figure}[ptbh]
\begin{center}
\begin{minipage}[h]{4.0in}
\epsfig{file=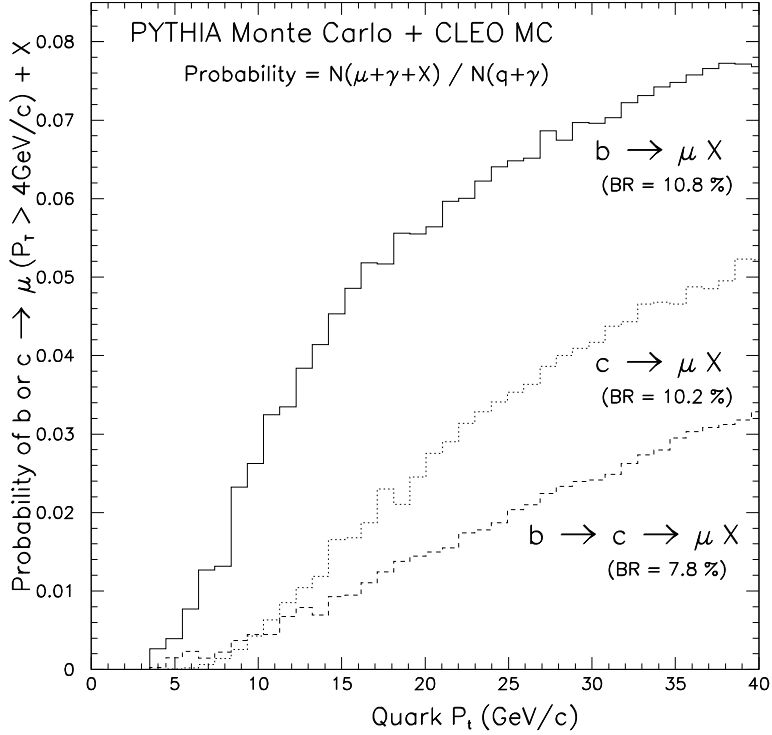,width=\linewidth}\caption
{The probability for producing a muon with $P_T >$4 GeV in the semi-leptonic decay of
a heavy quark is shown as a function of the quark $P_T$.  These were generated
using the PYTHIA Monte Carlo program to simulate the Compton \gm
process.  The number of muons produced in association with the photon, N($\gamma
+ \mu$ + X), includes the relevant branching ratios (shown in parentheses). The quark and muon pseudorapidity are required to be within $|{\eta
}%
|<1.0$.  The contribution from direct and sequential decays of the bottom quark are shown,  as
well as the direct decay of the charm quark.}%
\end{minipage}
\end{center}
\label{prob}\end{figure}

\begin{figure}[ptbh]
\begin{center}
\begin{minipage}[h]{4.0in}
\epsfig{file=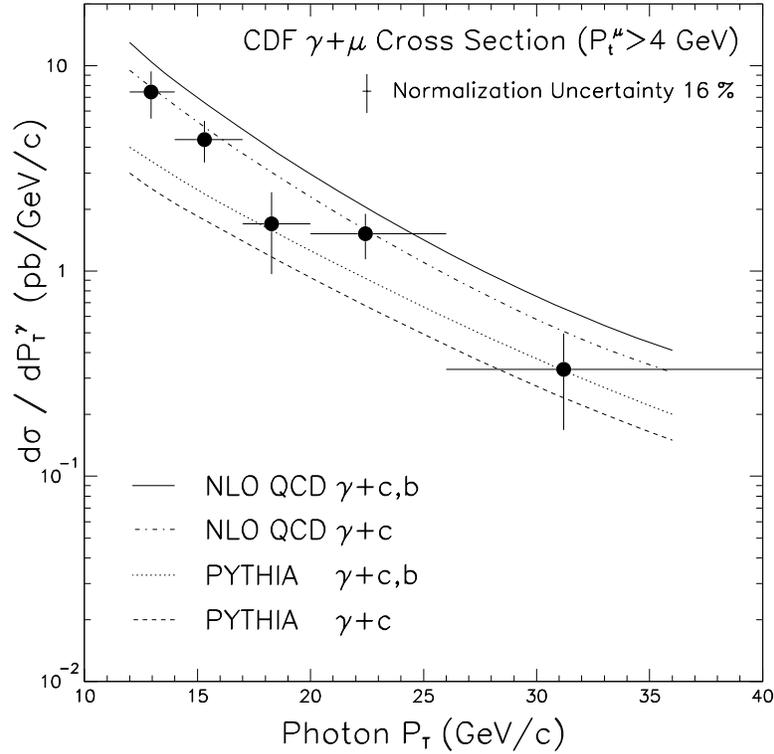,width=\linewidth}\caption{The measured \gm
cross section is shown as a function of photon $P_T$. The photon pseudorapidity
is required to be within $|{\eta}%
|<0.9$ and the muon pseudorapidity is required
to be within
$|{\eta}|<1.0$.  There is an overall 16\% normalization 
uncertainty (not shown) on the data.  The measurement is compared to QCD predictions from
the PYTHIA Monte Carlo as well as NLO QCD calculations.  Both 
the total contribution from  $\gamma+(b+c)$ and the individual 
$\gamma + c$ contribution are plotted.}%
\end{minipage}
\end{center}
\label{xsec}\end{figure}
\begin{figure}[ptbh]
\epsfig{file=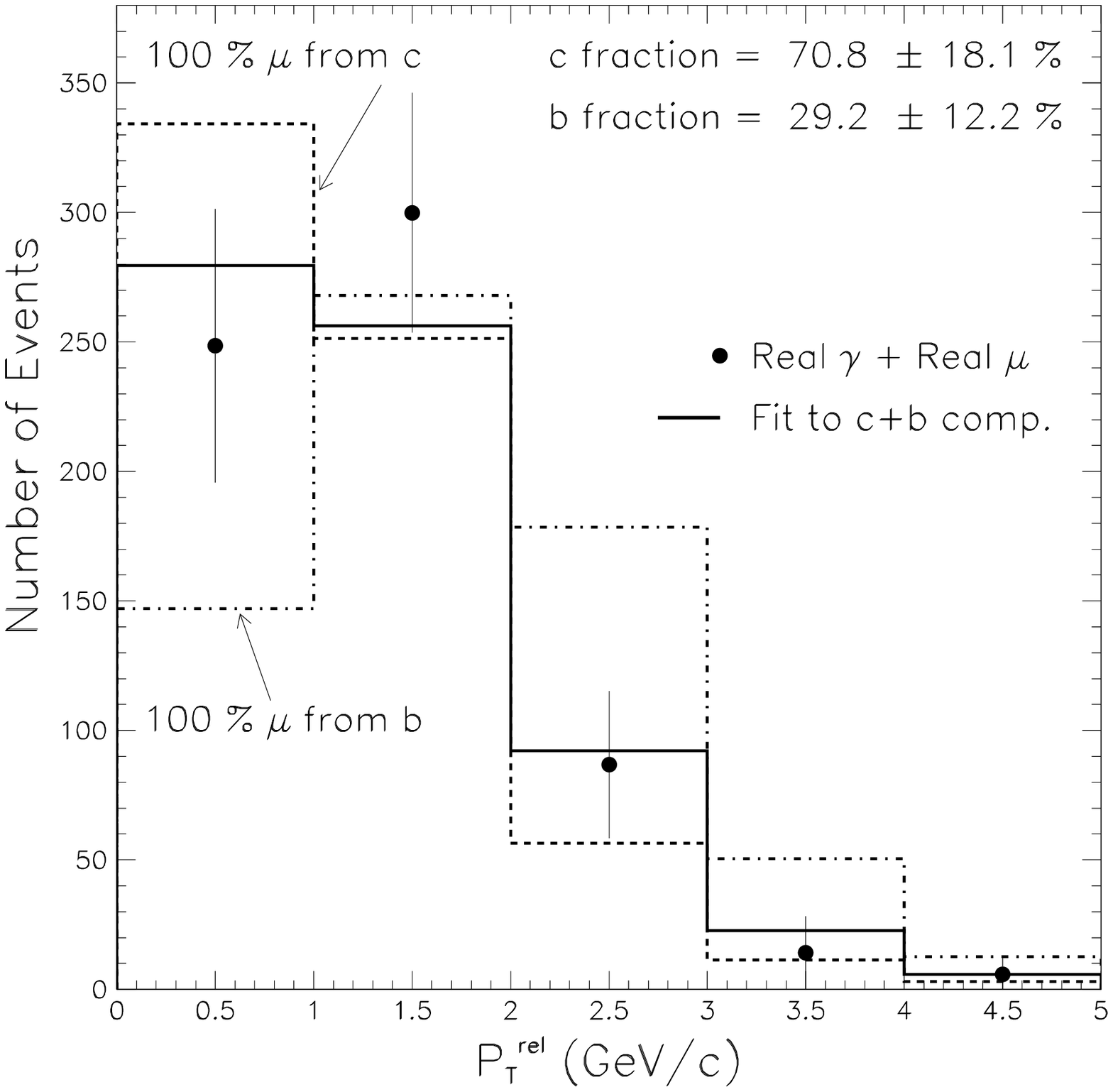,width=0.5\linewidth} \epsfig
{file=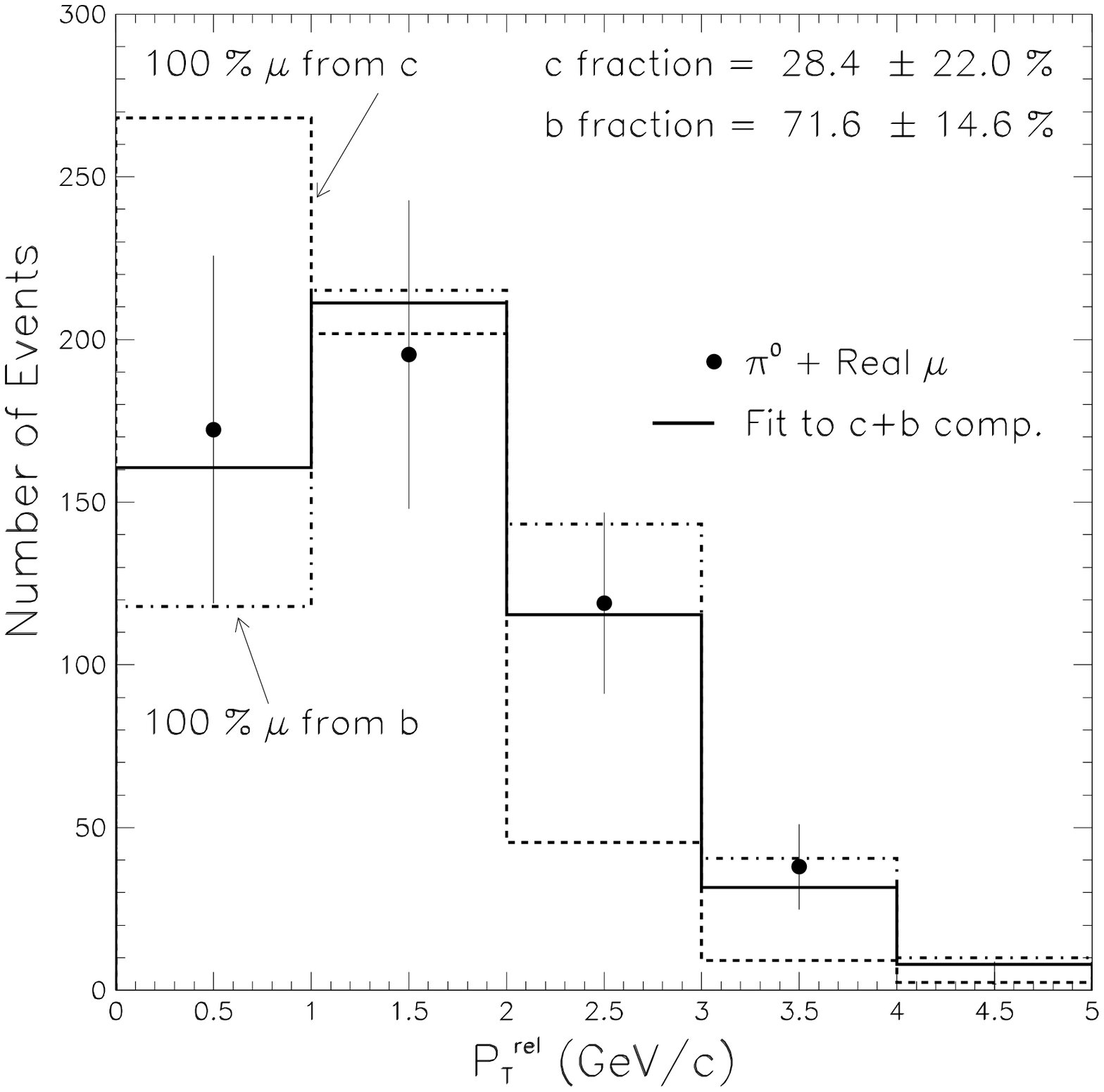,width=0.5\linewidth}\caption{The \ptrel method is used
to measure the sample composition of $\gamma$+heavy quark (left figure), and
$\pi^{0}$+heavy quark (right figure). In both figures, the points are the data after subtracting the contributions from fake muons. The solid curves are the best
fit to the data, the dashed curves are the expected \ptrel distribution of a
sample that is 100\% charm, and the dotted curves are the expected \ptrel
distribution of a sample that is 100\% bottom. }%
\label{phomucomp}%
\end{figure}
\end{document}